\documentclass[prl,twocolumn,showpacs]{revtex4}

\usepackage{graphicx}
\usepackage{dcolumn}
\usepackage{bm}

\begin{document}

\title{Half-filled 2D bilayers in a strong magnetic field: Revisiting the $\nu=1/2$ fractional 
quantum Hall effect}

\author{S. Das Sarma and Michael R. Peterson} \affiliation{Condensed
Matter Theory Center, Department of Physics, University of Maryland,
College Park, MD 20742}

\date{\today}

\begin{abstract}
We examine the quantum phase diagram of the fractional quantum Hall
effect in the lowest Landau level in half-filled bilayer structures as
a function of tunneling strength and layer separation.  Using
numerical exact diagonalization we investigate the important question
of whether this system supports a fractional quantum Hall effect
described by the non-Abelian Moore-Read Pfaffian state in the strong
tunneling regime. We find that, although it is in principle possible,
it is unlikely that the non-Abelian FQHE exists in the lowest Landau
level.  We establish that all so far observed FQHE states in
half-filled lowest Landau level bilayers are most likely described by
the Abelian Halperin 331 state.
\end{abstract}

\pacs{73.43.-f, 71.10.Pm}

\maketitle

Two important developments have rekindled interest in the
phenomena of even-denominator incompressible fractional quantum Hall
(FQH) states in bilayer semiconductor structures.  The first is the
recent intriguing experimental observation by Luhman \textit{et
al}.~\cite{luhman} of two distinct even-denominator (filling factors $\nu=1/2$ and
1/4) FQH states in a very wide ($\sim 600 \AA$) single quantum
well (WQW) structure at very high ($>40 T$) magnetic fields.  The
second development, motivated by implications for fault-tolerant
topological quantum computation~\cite{sds-tqc}, is the great deal of recent
theoretical and experimental interest in the
possible non-Abelian nature of the $\nu=5/2$ second Landau level (SLL)
FQH state in the single-layer system~\cite{tqc-rmp}.  This leads to the
interesting as well as important question whether a non-Abelian
$\nu=1/2$ FQH state, i.e., the analog of the possibly non-Abelian
$\nu=5/2=2+1/2$ SLL single-layer state, can exist in the lowest Landau
level (LLL) under experimentally observable conditions.  This question
has a long history~\cite{LLLhistory} in the theoretical literature going back
to the early 1990s.  Motivated by these considerations we revisit the
$\nu=1/2$ FQHE in bilayer structures (with each layer having $\nu=1/4$
on average) by theoretically investigating the quantum phase diagram
for the LLL $\nu=1/2$ bilayer FQH state using the FQH spherical system
finite size exact diagonalization technique.  We obtain an approximate
quantum phase diagram for the spin-polarized $\nu=1/2$ bilayer FQH
system in the LLL in the parameter space of inter-layer separation
($d$), inter-layer tunneling strength ($t$) or, equivalently, the
symmetric-antisymmetric gap energy, and the width or
thickness ($w$) of the individual wells (for simplicity, we assume the
two wells to be identical).  We take the system to be fully
spin-polarized since all experimental $\nu=1/2$ FQH states appear to
be fully spin-polarized.

Our main findings are: (i) the recently observed WQW $\nu=1/2$
FQHEs~\cite{luhman,shayegan-new} are strong-pairing Abelian Halperin
331 FQH states~\cite{halperin-331} which, however, sit close to the
boundary between the Abelian 331 and the weak-pairing non-Abelian
Pfaffian (Pf) FQH state~\cite{mr-pf}, and (ii) it may be conceivable,
as a matter of principle, to realize the LLL $\nu=1/2$ Pfaffian
non-Abelian FQHE in very thick bilayers, but as a matter of practice,
this is unlikely since the $\nu=1/2$ FQHE gap is extremely
small (perhaps zero) in the parameter regime where the Pf is more
stable than the 331 phase.  Our findings about the fragility of the
LLL $\nu=1/2$ non-Abelian Pf state are consistent with recent
conclusions~\cite{recent-theory,storni}, but our main focus in the
current work is in understanding the $\nu=1/2$ bilayer quantum phase
diagram treating $t$, $d$, and $w$ as independent tuning parameters of
the system Hamiltonian.

Before presenting our results, we discuss the context of our
theoretical investigation.  The proposed Moore-Read Pf state is a
weak-pairing single-layer FQH state at $\nu=1/2$ which, in principle,
applies to any orbital LL (i.e., LLL as well as SLL).  Thus, as
a matter of principle the $\nu=1/2$ LLL single-layer Pf FQHE is
certainly a possibility although it has never been observed
experimentally.  The best existing numerical
work~\cite{recent-theory,storni} indicates that either the
single-layer $\nu=1/2$ LLL Pf state does not exist in nature or if it exists,
does so only in rather thick 2D layers with an extremely
small FQH excitation gap, making it impossible or very difficult to
observe experimentally.  By contrast, the single-layer $\nu=5/2$ SLL
FQHE is observed routinely, albeit at low temperatures ($\lesssim 100
mK$), in high mobility ($\gtrsim 10^7 cm^2/V s$) samples, and with a
rather small (but experimentally accessible) activation gap ($\sim
100$-$500 mK$).  In fact, it has been pointed out that the
experimental $\nu=5/2$ FQHE is always among the strongest (along with
the $\nu=7/3$ and 8/3 FQHE) observed FQH states in the SLL.

Instead of studying a single-layer 2D system we
concentrate on the spin-polarized bilayer system assuming an
arbitrary tunneling strength $t$ (proportional to the
symmetric-antisymmetric gap) and an arbitrary layer separation $d$.
We numerically obtain the quantum phase diagram at $\nu=1/2$ for this
bilayer system in the $t$-$d$ space, concentrating entirely on the
Pfaffian and the 331 FQHE phases.  Our focus on the bilayer
spin-polarized $\nu=1/2$ FQHE is consistent with the
experimental fact that the incompressible $\nu=1/2$ FQHE has so far
been observed only in effective bilayer structures.  The goal is to
carry out an extensive comparison with all existing bilayer $\nu=1/2$
FQH experimental observations to ascertain any hint of a Pfaffian
$\nu=1/2$ state for large values of $t$.  It is obvious that our model
system is an effective bilayer (single-layer) $\nu=1/2$ system for
small (large) values of $t$, and therefore by studying 
the quantum phase diagram as a function of $t$ (and $d$) we hope to
shed light on the possible existence of a single-layer $\nu=1/2$ FQHE
in real systems.  Recent FQHE experiments at $\nu=1/2$ for large
values of the tunneling strength make it imperative that a
theoretical analysis be carried out to achieve a proper
qualitative understanding of the experimental situation~\cite{luhman,shayegan-new}.

We use a simple model Hamiltonian incorporating both 
finite tunneling and  finite layer separation for our bilayer FQH
system:
\begin{eqnarray}
\hat{H}&=&\sum_{i<j}^N [V_\mathrm{intra}(|\mathbf{r}_i-\mathbf{r}_j|) + 
V_\mathrm{intra}(|\tilde{\mathbf{r}}_i-\tilde{\mathbf{r}}_j|)\nonumber \\
&+& V_\mathrm{inter}(|\mathbf{r}_i-\tilde{\mathbf{r}}_j|)] - t \hat{S}_x \;,
\label{eq-1}
\end{eqnarray}
where $\mathbf{r}_i$($\tilde{\mathbf{r}}_i$) is the position of the
$i$-th electron in the right(left) layer.  In Eq.~(\ref{eq-1}),
$V_\mathrm{intra}(r)=e^2/(\kappa\sqrt{r^2+w^2})$ and
$V_\mathrm{inter}(r)=e^2/(\kappa\sqrt{r^2+d^2})$ are the intralayer
and interlayer Coulomb interaction incorporating a finite layer
width $w$ and a center-to-center interlayer separation $d$ ($>w$ by
definition).  The $x$-component of the pseudo-spin operator
$\hat{S}_x$ controls the tunneling between the two quantum wells with
large $t$ denoting strong tunneling.  We numerically diagonalize
$\hat{H}$ in the spherical geometry assuming specific values of $w$,
$d$, and $t$ (each expressed throughout in dimensionless units
using the magnetic length $l=(c\hbar/eB)^{1/2}$ as the length unit
and the Coulomb energy $e^2/(\kappa l)$, where $\kappa$ is the
background dielectric constant, as the energy unit).
Following the standard well-tested procedures~\cite{LLLhistory,recent-theory,storni} used extensively in the
FQHE  literature, we calculate the overlap between the
exact numerical $N$ electron ground state wavefunction of the Coulomb Hamiltonian
defined by Eq.~(\ref{eq-1}) and the candidate $N$ electron variational states which
are the Halperin 331 strong-pairing~\cite{halperin-331} and the Moore-Read Pfaffian
weak-pairing~\cite{mr-pf} wavefunctions:
\begin{eqnarray}
\Psi_{331}&=&\prod_{i<j}^{N/2}(z_i-z_j)^3\prod_{i<j}^{N/2}(\tilde{z}_i-\tilde{z}_j)^3
\prod_{i,j}^N (z_i-\tilde{z}_j)\\
\Psi_\mathrm{Pf}&=&\mathrm{Pf}\left\{\frac{1}{z_i-z_j}\right\}\prod_{i<j}^N(z_i-z_j)^2  \;,
\end{eqnarray}
respectively, where $z=x-iy$ is the electron coordinate in the plane.  
We ensure that (i) the ground state is homogeneous, i.e., has zero
total orbital angular momentum, and (ii) there is a gap, the FQH
excitation gap, separating the ground state from all excited
states.  The results shown in this paper all use an $N=8$ electron
system, but larger systems show the same qualitative
features.  
Since the theoretical techniques are standard, we do not
give the details, concentrating instead on the results and their
implications for $\nu=1/2$ FQHE experiments.

The calculated overlap and gap determine the nature of the FQHE and
its strength at $\nu=1/2$ in our theory.  We operationally define the
system to be in the 331 (Pf) phase depending on the overlap
with the 331 (Pf) state being the larger of the two.  We emphasize that
our work is a comparison between these two incompressible states only,
and we cannot comment on the possibility of some other state
(i.e., neither 331 nor Pf) being the ground state.  We
do, however, believe that if the system is incompressible at a
particular set of parameter values (i.e., $d$, $t$,
$w$, etc.), it is very likely to be described by one of
these two candidate states, 331 or Pf, since no
other proposed incompressible state exists in the literature for
$\nu=1/2$ spin-polarized bilayer or single-layer LLL systems.  We
cannot, however, rule out the possibility that the real system has a
compressible ground state (without manifesting FQHE), e.g., a
composite fermion sea, not considered in our calculation.

\begin{figure*}[t]
\begin{center}
\mbox{\includegraphics[width=5.9cm,angle=0]{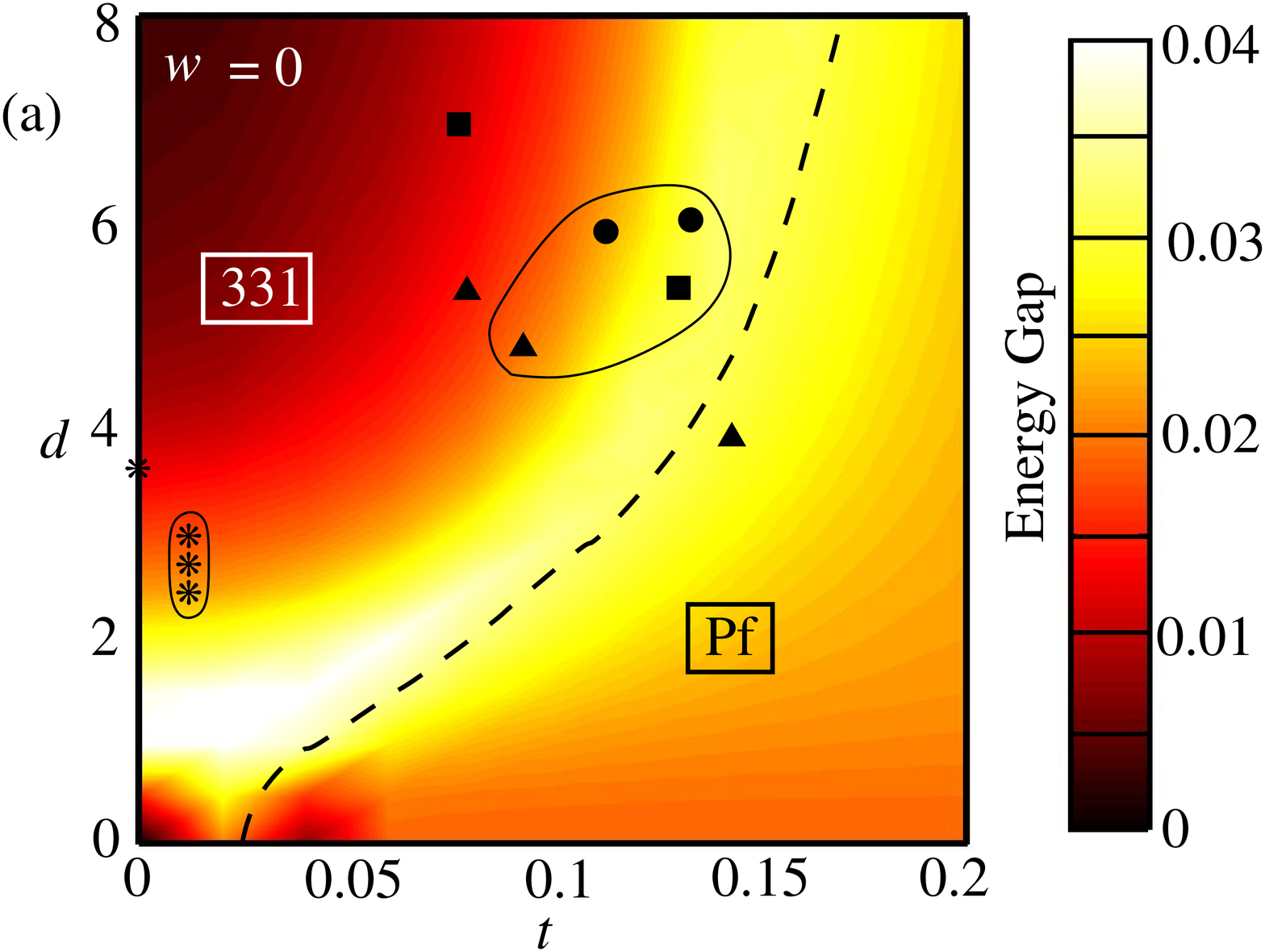}}
\mbox{\includegraphics[width=5.9cm,angle=0]{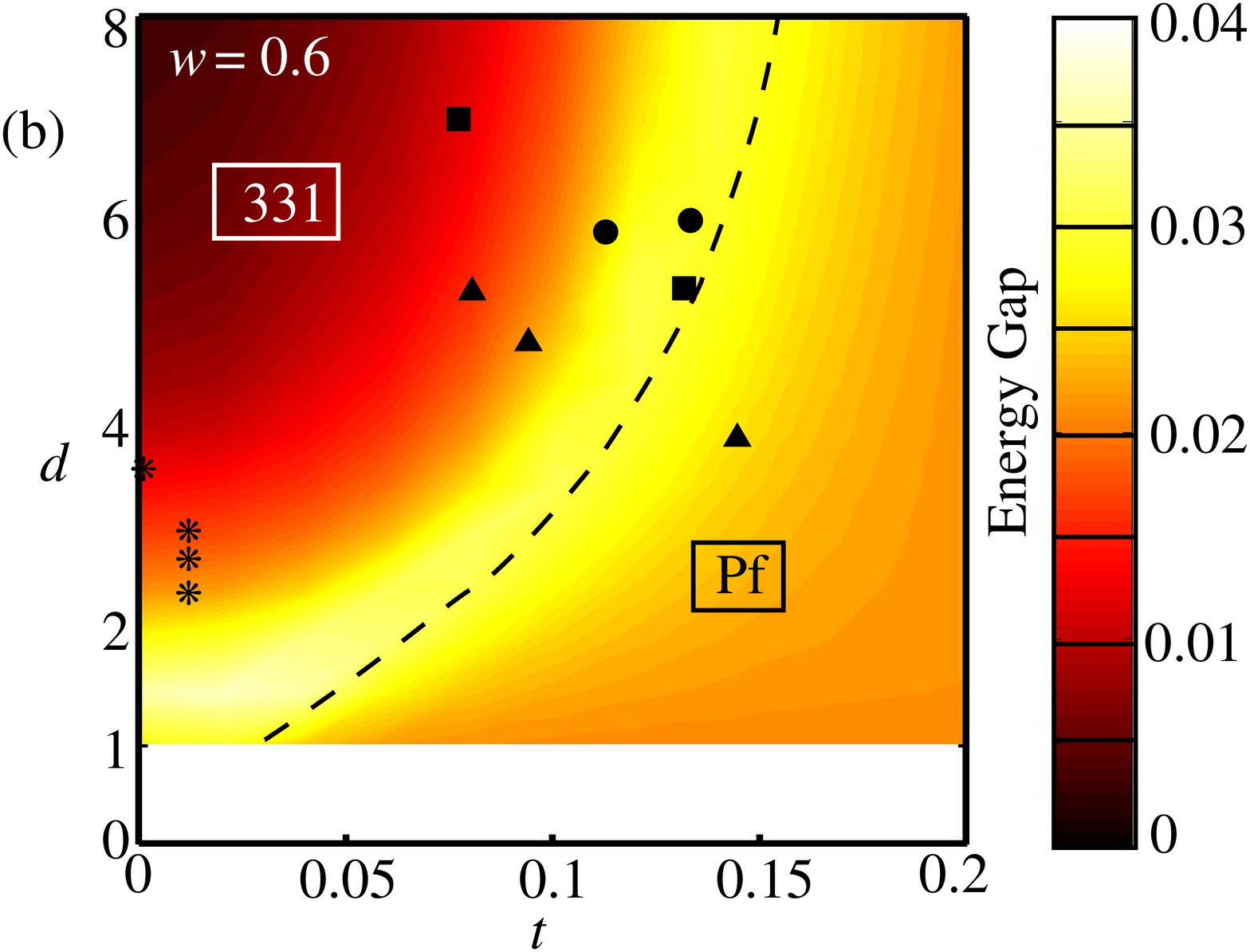}}
\mbox{\includegraphics[width=5.9cm,angle=0]{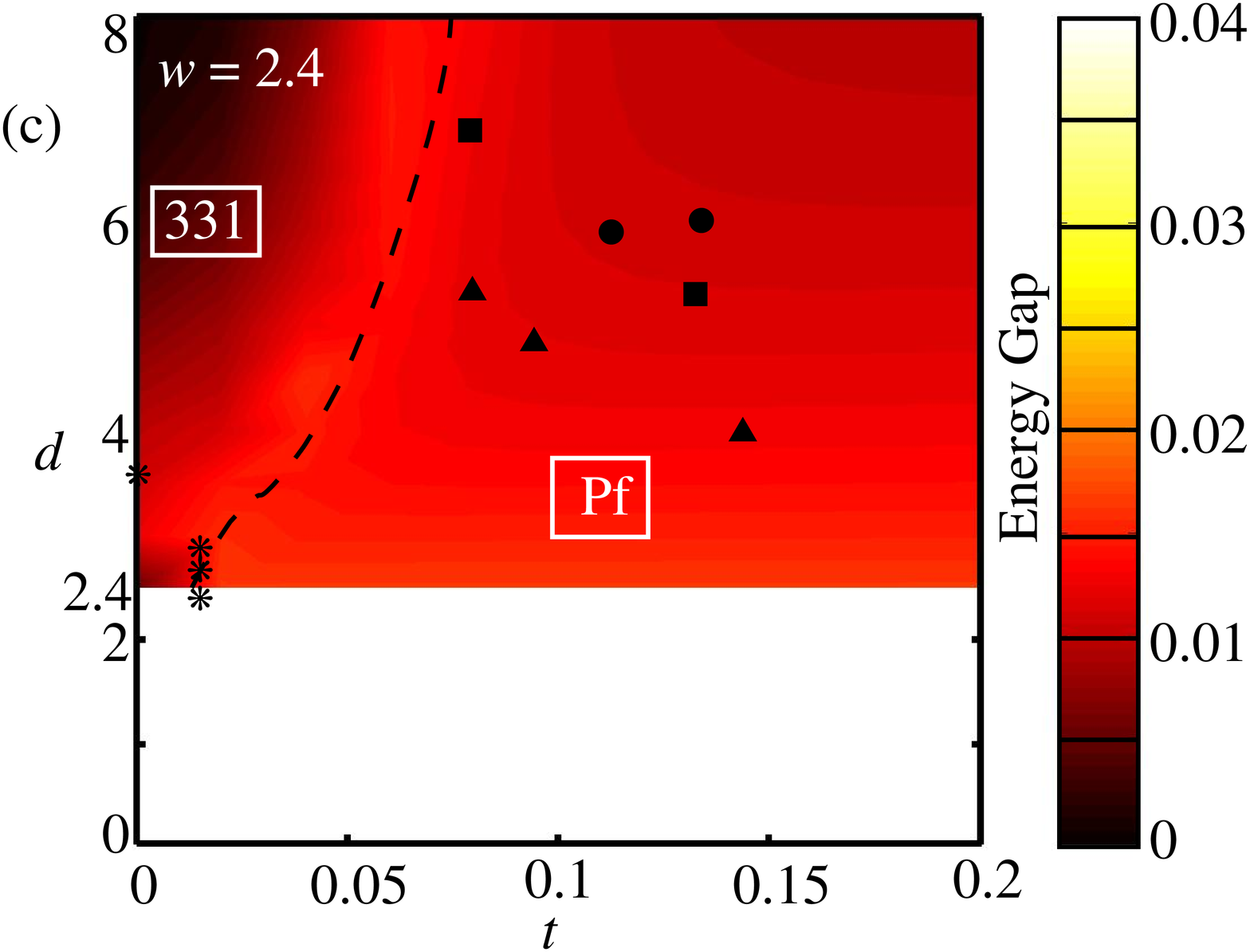}}
\end{center}
\caption{(color online) Quantum phase diagram (QPD) and FQHE gap
(color coded) versus layer separation $d$ and tunneling strength $t$
for widths (a) $w=0$, (b) $w=0.6$, and (c) $w=2.4$.  For the QPD, the
331 and Pf phases (as discussed in the text) are separated by a dashed
black line and labeled appropriately.  The FQHE gap is given as a
contour plot with color coding given by the color-bar from dark to
light, i.e., white being a largest value of 0.4 and black being value
of 0.  The asterisks, triangles, circles, and squares correspond to
the different experiments in
Refs.~\onlinecite{eisenstein},~\onlinecite{shayegan-old},~\onlinecite{shayegan-new}
and~\onlinecite{luhman}, respectively.  Only experimental points
showing FQHE are within the large solid circles in (a) with the lower
smaller (upper larger) circles indicating experiments in
double-quantum-well (single-WQW) structures.  We note that the single
triangle on the Pf side of the QPD does not manifest any experimental
FQHE indicating that the theoretical gap may be overestimated for the
Pf state.  It is obvious from the figures that the Pf state becomes
more dominant, albeit with very small FQH excitation gap, for larger
values of $w$.}
\label{fig1}
\end{figure*}

We first show in Fig.~\ref{fig1} our numerically calculated FQHE quantum phase
diagram (QPD) for the bilayer $\nu=1/2$ system in the $t$-$d$ space
with the color coding indicating the numerical FQH gap strength and
the dashed line separating the 331 phase from the Pf phase (i.e., the
overlap with 331 (Pf) being larger above (below) the dashed line).
Note that the dashed line is only an operational phase boundary within
our calculation since all we know is that the 331 (Pf) has higher
(lower) overlap above (below) this line.  The calculated overlap for
each phase, i.e., 331 (Pf) above (below) the dashed line, varies
between $\sim0.6$ and $\sim0.96$ for our 8-particle system.  We show
the QPD for three values of the layer width parameter (a) $w=0$; (b)
$w=0.6$; (c) $w=2.4$.  The zero- (Fig.~\ref{fig1}(a)) and the
intermediate-width (Fig.~\ref{fig1}(b)) results are of physical
relevance whereas the (unrealistically) large width results
(Fig.~\ref{fig1}(c)) are provided here only for completeness (since
this is the regime where the Pf state dominates over the 331 state in
the QDP).  We note that we are using the simplistic Zhang-Das Sarma
(ZDS) model~\cite{zds} for describing the well width effect, and
crudely speaking $w=1$ in the ZDS model corresponds roughly to
$w_{QW}\approx 6$ where $w_{QW}$ is the corresponding physical quantum
well width.  For a single WQW, where the effective bilayer is created
by the self-consistent potential of the electrons themselves, our
model $w$ is typically much less than the total width $W$ of the
WQW--very roughly speaking $w\sim W/6$, and $d\sim W/2$.  As
emphasized above, we treat $t$, $d$, and $w(<d)$ as independent tuning
parameters.

In Fig.~\ref{fig1}, we have put as discrete symbols all existing
$\nu=1/2$ bilayer experimental data (both for double quantum
well systems and single wide quantum wells) in the literature,
extracting the relevant parameter values (i.e., $d$ and $t$) from the
experimental works~\cite{eisenstein,shayegan-old,luhman,shayegan-new}.
Because of the ambiguity and uncertainty in the definition of $w$, we
have put the data points on all three QPDs shown in Fig.~\ref{fig1}
although the actual experimental width values correspond to only
Figs.~\ref{fig1}(a) and (b).

Results shown in Fig.~\ref{fig1} bring out several important points of
physics not clearly appreciated earlier in spite of a great deal of
theoretical exact diagonalization work on $\nu=1/2$ bilayer FQHE: 
(i) It is obvious that large (small) $t$
and small (large) $d$, in general, lead to a decisive preference for
the existence of $\nu=1/2$ Pf (331) FQHE.  The fact that large $t$
values would preferentially lead to the Pf state over the 331 state
is, of course, expected since the system becomes an effective
one-component system for large tunneling strength.  (ii) What is,
however, not obvious, but apparent from the QPDs shown in
Fig.~\ref{fig1}, is that the FQH gap (given in color coding in the
figures) is maximum near the phase boundary between  331 and Pf.  
(iii) Another non-obvious result is the persistence of the 331
state for very large (essentially arbitrarily large!)  values of the
tunneling strength $t$ as long as the layer separation $d$ is also
large--thus having a large $t$ by itself, as achieved in the Luhman
\textit{et al}. experiments~\cite{luhman}, is not enough to realize
the single-layer $\nu=1/2$ Pf FQHE, one must also have a relatively
small value of layer separation $d$ so that one is below the phase
boundary (dashed line) in Fig.~\ref{fig1}.  The explanation for the
Luhman experimental $\nu=1/2$ FQHE being a 331 sate, as can be seen in
Fig.~\ref{fig1}, is indeed the fact that both $t$ and $d$ are large in
these samples making 331 a good variational state.  (iv) An important
aspect of Fig.~\ref{fig1} is that the Pf FQHE gap tends to be very
small--this is particularly true for larger values of $w$, where the
Pf overlap is larger.  This implies, as emphasized by Storni
\textit{et al}.~\cite{storni}, that the observation of a $\nu=1/2$ Pf
state is unlikely since the activation
gap would be extremely (perhaps even vanishingly) small.  (v) For
larger values of $w$ (and large $t$), our calculated QPD is dominated
by the Pf state--particularly for the unrealistically large width
$w=2.4$ (corresponding to $w_{QW}\sim 14$!) where all the experimental
$d$ and $t$ values fall in the Pf regime of the phase diagram.  We
emphasize, however, that this Pf-dominated large-$w$ (and large-$t$)
regime will be difficult (perhaps even impossible) to access
experimentally since the FQH gap would be apparently extremely small 
as in Fig.~\ref{fig1}.

In discussing Fig.~\ref{fig1} further, we mention that our 331 (Pf)
regimes not only have the wavefunction overlap with the corresponding
331 (Pf) state being larger than the other, but also the
calculated expectation value $\langle \hat{S}_x\rangle\approx 0$ (4) in the
331 (Pf) regime.  Thus, our QPD is consistent with both the overlap
and the pseudo-spin calculation as obtained from exact
diagonalization.

We now discuss the published experimental results in light of our
theoretical QPD.  First, we note that most of the existing
experimental points fall on the 331 side of the phase diagram which is
consistent with our QPD in Fig.~\ref{fig1}.  In particular, only 
samples on the 331 side of the QDP with reasonably large FQH
gaps, i.e., the data points close to the phase boundary, exhibit
experimental FQHE.  By contrast, the one data point (in
Figs.~\ref{fig1}(a) and (b)) on the Pf side of the phase boundary does
not manifest any observable FQHE in spite of its location being in a
regime of reasonable FQH excitation gap according to our phase
diagram.  This is consistent with the finding of Storni \textit{et
al.}~\cite{storni} that the $\nu=1/2$ FQH Pf gap in a single-layer
system is likely to be vanishingly small in the thermodynamic limit.
It is, therefore, possible that much of the Pf regime in our QPD has a
much smaller excitation gap than what we obtain on the basis of our
$N=8$ particle diagonalization calculation.  We refer to Storni
\textit{et al.}~\cite{storni} for more details on the theoretical
status of the single-layer LLL $\nu=1/2$ FQHE.

\begin{figure}[]
\begin{center}
\mbox{\includegraphics[width=4.5cm,angle=-90]{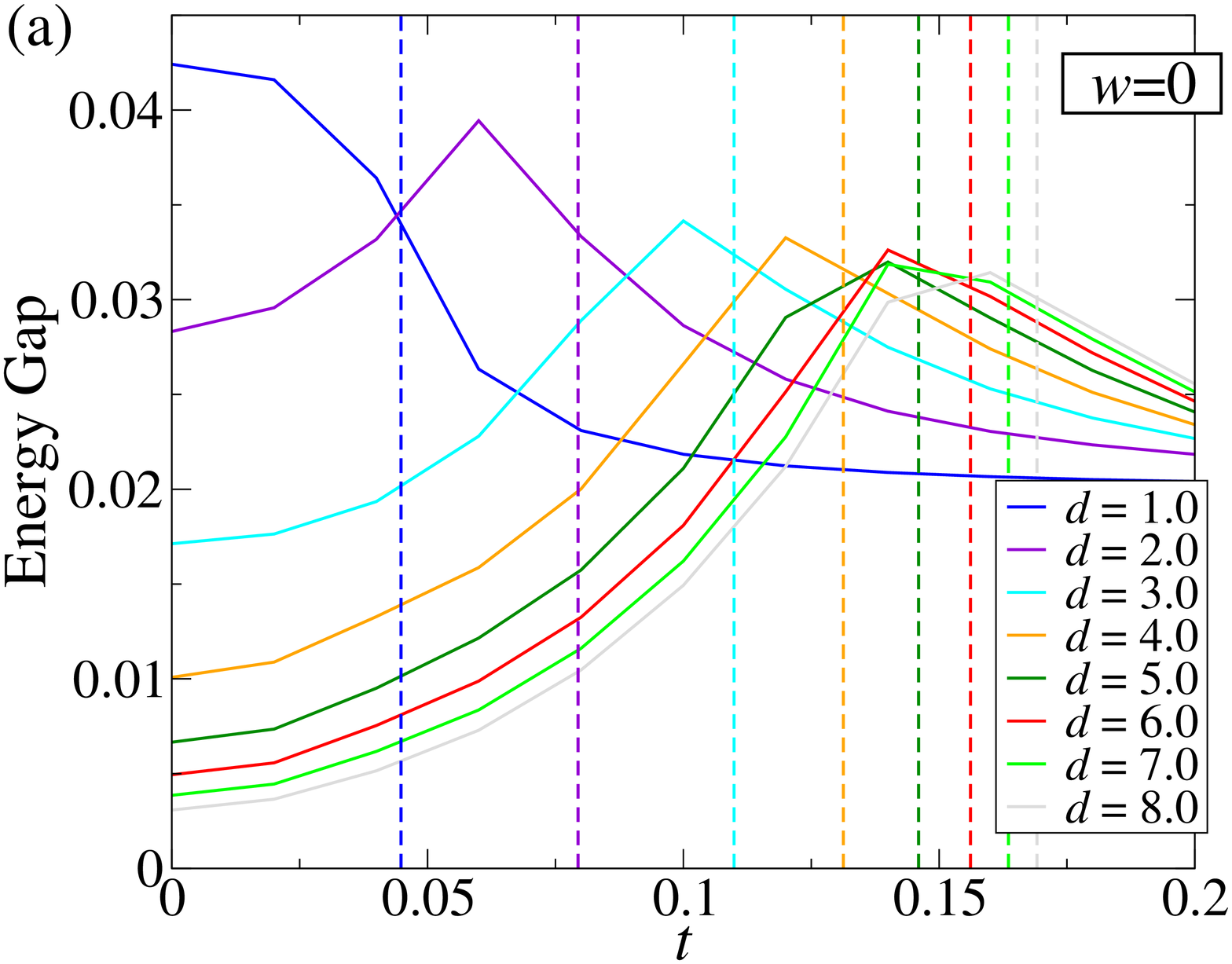}}
\mbox{\includegraphics[width=4.5cm,angle=-90]{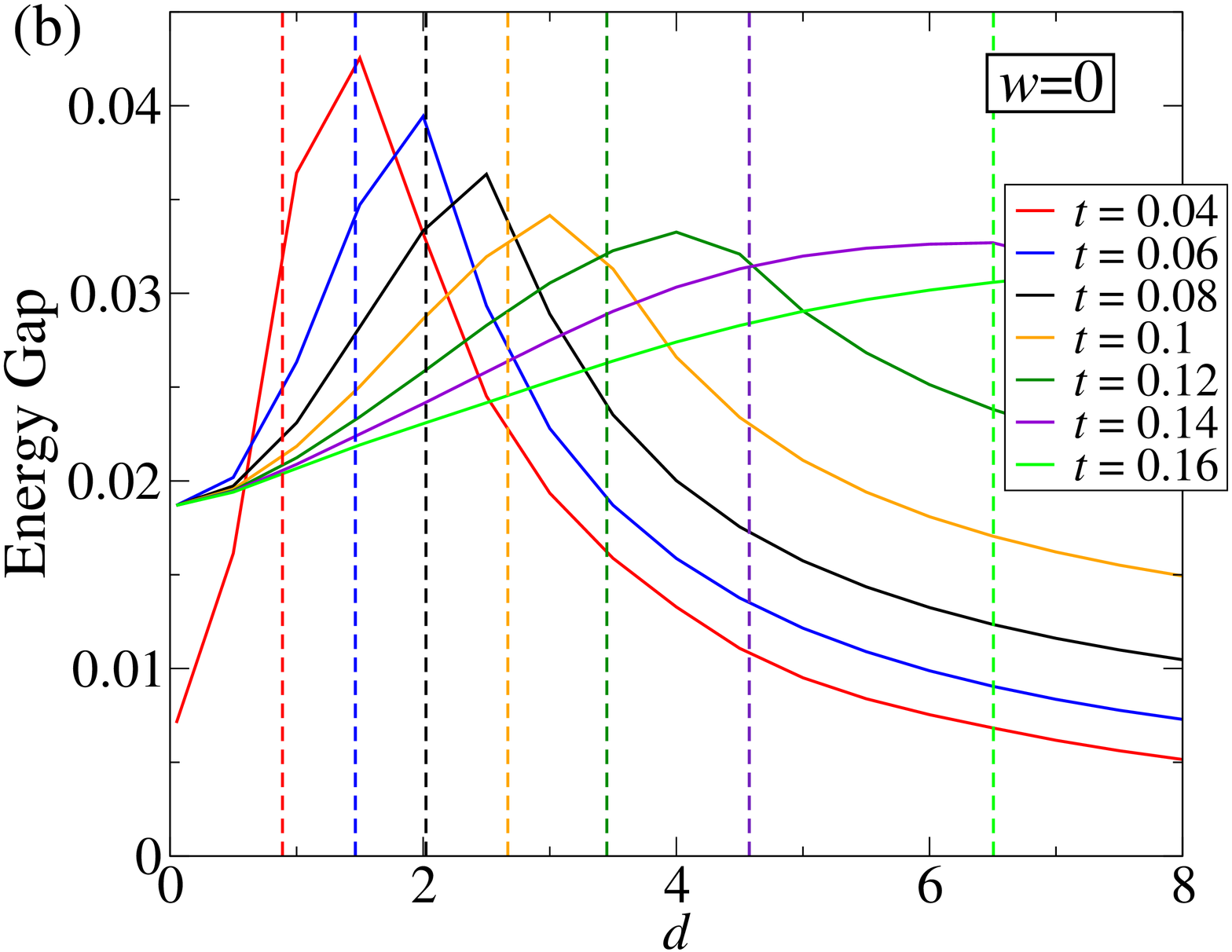}}
\end{center}
\caption{(color online) FQHE energy gap versus (a) tunneling strength $t$
for a few values of layer separation $d$ and (b) layer separation $d$
for a few values of $t$ (both consider zero width $w=0$).  A dashed
vertical line of the same shade corresponds to the boundary
between the Pfaffian phase (right/left of the line for (a)/(b))
and the 331 phase (left/right of the line for (a)/(b)).}
\label{fig2}
\end{figure}

For a more detailed view of the $\nu=1/2$ bilayer FQHE, we show in
Figs.~\ref{fig2}(a) and (b), respectively, our calculated FQHE gap
as a function of $t$ (for a few fixed $d$ values) and as a function of
$d$ (for a few fixed $t$ values).  In each figure, we also depict the
line separating the 331 (smaller $t$/larger $d$) and the Pf (larger
$t$/smaller $d$) regimes in the phase diagram.  The qualitatively
interesting point is, of course, the non-monotonicity in the FQHE
gap as a function of $t$ or $d$ with a maximum close (but always on
the 331 side) to the phase boundary.  The non-monotonicity in the FQHE
gap as a function of $t$ (but not $d$) was earlier pointed out, but
our finding that the peak lies \textit{always} on the 331 side of the
phase boundary is a new result.  We emphasize that the FQHE gap peak
lying always on the 331 side of the phase boundary is strong evidence
that the 331 phase is the dominant FQH phase in $\nu=1/2$
systems.  We believe that the only chance of observing the $\nu=1/2$
Pf FQHE is to look on the Pf side of phase boundary at
fairly large values of $d$ and $t$.  This is in sharp contrast to the
SLL $\nu=5/2$ bilayer FQHE where we recently showed that there are two
sharp ridges far away from each other in the $d$-$t$ space
corresponding to the $\nu=5/2$ Pf and 331 bilayer
phases~\cite{mrp-sds-sll-bilayer}.  We note that for unrealistically
large $w$ (Fig.~\ref{fig1}(c)), Pf dominates over 331 but the FQHE gap
becomes extremely small everywhere.

We conclude by commenting on the nature of the quantum phase
transition (QPT) between the 331 and the Pf phase in the $d$-$t$
space.  It may appear at first sight that our work implies a
continuous QPT from the strong-pairing 331 to the weak-pairing Pf
state with increasing (decreasing) $t$ ($d$).  This is, however,
deceptive since all we are doing is \textit{comparing these two
phases} using a finite size diagonalization study for discrete values
of $t$ and $d$.  It is entirely possible that a completely different
phase, e.g. a compressible composite fermion Fermi liquid phase, has
lower energy and intervenes between the 331 and Pf phases so that the
system goes from 331 to Pf (or vice versa) through two first-order
transitions.  There is independent numerical evidence that the
compressible composite fermion sea indeed has a lower ground state
energy than the Pf state in a single-layer $\nu=1/2$ (but not 5/2)
system which corresponds to the large $t$ (and small $d$) regime in
our QPD.  This would indicate first order transitions in going from
331 to the Pf (if it exists) through the compressible phase.  What we
have shown here is that if the $\nu=1/2$ bilayer Pf phase exists at
all, then it would manifest most strongly in wide samples and close to
the phase boundary with the 331 phase, but will have an extremely
small FQHE excitation gap.

This work is supported by Microsoft Q and DARPA QuEST.

\end{document}